\documentclass[11pt]{article}
\usepackage{geometry}                
\usepackage{graphicx}
\usepackage{amssymb}
\usepackage{multirow}
\usepackage{amsmath}
\usepackage{epstopdf}
\usepackage{authblk}
\usepackage{tikz}
\usepackage{amsthm}
\usepackage{colortbl}
\usepackage{amsmath}
\newtheorem{theorem}{Theorem}

\newtheorem{conjecture}{Conjecture}

\usepackage[all,cmtip]{xy}
\title{Three facets of mathematical cancer biology research}
\author[1]{Yue Wang}
\affil[1]{Department of Computational Medicine, University of California, Los Angeles, California, United States of America}

\date{}                                           

\begin{document}
\maketitle

\begin{abstract}
Cancer, as the uncontrollable cell growth, is related to many branches of biology. In this review, we will discuss three mathematical approaches for studying cancer biology: population dynamics, gene regulation, and developmental biology. If we understand all biochemical mechanisms of cancer cells, we can directly calculate how the cancer cell population behaves. Inversely, just from the cell count data, we can use population dynamics to infer the mechanisms. Cancer cells emerge from certain genetic mutations, which affect the expression of other genes through gene regulation. Therefore, knowledge of gene regulation can help with cancer prevention and treatment. Developmental biology studies acquisition and maintenance of normal cellular function, which is inspiring to cancer biology in the opposite direction. Besides, cancer cells implanted into an embryo can differentiate into normal tissues, which provides a possible approach of curing cancer. This review illustrates the role of mathematics in these three fields: what mathematical models are used, what data analysis tools are applied, and what mathematical theorems need to be proved. We hope that applied mathematicians and even pure mathematicians can find meaningful mathematical problems related to cancer biology.
\end{abstract}

\section{Introduction}
Cancer, as a group of more than 100 types of diseases, is one of the main causes of human death. It is the dream of countless scientists to further understand cancer and finally cure cancer. However, many experiments might cause harm \cite{li2021chronic}, and cannot be performed on human beings. Therefore, given limited types of data, we have to develop indirect methods to study cancer biology, especially with mathematical tools.

In this review, we will go through some mathematical biology studies in these fields, and discuss how research in these fields improves our understanding of cancer biology. This review is targeted to researchers who have backgrounds in mathematics and want to contribute to cancer biology research. We will discuss what mathematical models are used, what data analysis tools are applied, and what mathematical theorems need to be proved.

Population dynamics studies how the population level changes along time under different circumstances. The growth of tumor can be described by population dynamics. In Section~\ref{sec2}, we introduce some mathematical studies in cancer cell population dynamics in normal environment and under treatment. These studies illustrate what mathematical problems arise in population dynamics, and what insights can be obtained from mathematical analyses of population data. Specifically, there are still open problems arising from related stochastic models.

Gene regulation describes how the expression of one gene affects the expression of another gene. Cancer-related genetic mutations can deviate the normal cell fate through gene regulation. Nevertheless, gene regulation is difficult to determine directly by experiments. In Section~\ref{sec3}, we introduce some mathematical studies in the inference of gene regulation (both mutual regulation and autoregulation) from certain types of data. These studies represent a broad field of developing mathematical inference methods in gene regulation and other subjects. Various problems in probability and discrete mathematics need to be solved.

Developmental biology studies how organized normal tissues arise from a single zygote. It is the opposite of cancer biology, which studies the emergence of abnormal tissues. A developing embryo can invert the fate of cancer cells. In Section~\ref{sec4}, we introduce some mathematical studies in developmental biology, which could provide insights of studying cancer biology from the opposite direction. Specifically, we introduce an inference method for certain experiments, which leads to a coloring problem in lattice.

There have been many papers and books discussing the role of mathematics in cancer biology \cite{longo2018information,wodarz2005computational,altrock2015mathematics,gatenby2003mathematical,anderson2013mathematics}. We hope that this review can provide a little more insights and encourage applied mathematicians and even pure mathematicians to work on interesting problems related to cancer biology and contribute to the cure of cancer.

\section{Population dynamics}
\label{sec2}
The most significant feature of cancer cells is the uncontrollable growth. Therefore, measuring and studying the growth of cancer cell population should be a central topic in cancer research. Many studies build models for cancer cells living in a free environment \cite{jiang2017phenotypic}. Some other studies consider the effect of nutrients \cite{jacobs2022tumor} and drugs \cite{angelini2022model}.

Traditional population dynamics models are deterministic, based on ordinary differential equations (ODEs) \cite{zhou2014multi} or partial differential equations (PDEs) \cite{wang2022modelling}. Since cell growth is highly stochastic, various stochastic processes are used to model cellular clonal growth and evolution: birth-death processes \cite{wang2018some}, Poisson processes \cite{bartoszynski1981some}, Markov chains \cite{gupta2011stochastic}, random sums of birth-death processes \cite{dewanji2005generalized}, and branching processes \cite{jiang2017phenotypic}. Other stochastic approaches, such as random dynamical systems \cite{ye2016stochastic}, lifted Markov chains \cite{wang2020mathematical}, continuous Markov processes \cite{qian2020kinematic,yang2021potentials}, or more complicated models based on the above methods \cite{zhou2021dissecting}, also have the potential of modeling cell population growth.
\subsection{Phenotypic equilibrium phenomena}

Mathematical models based on different biological assumptions can produce different predictions, which can be verified by experiments. For instance, in the traditional hierarchical model \cite{reya2001stem,jordan2006cancer,dalerba2007cancer}, the transition from cancer stem cells to non-stem cancer cells is irreversible. In recent studies, there is evidence indicating that non-stem cancer cells can convert back to cancer stem cells in certain cancer types \cite{meyer2009dynamic,chaffer2013poised,quintana2010phenotypic,yang2012dynamic,fessler2015endothelial}. In mathematical models that consider such reversible transitions, there are two phenomena that do not exist in the hierarchical model: overshoot and phenotypic equilibrium \cite{niu2015phenotypic,chen2016overshoot}. Overshoot means that if the initial population value is lower than the steady state, the population might first increase above the steady state and then return to it. \textbf{Phenotypic equilibrium} means that starting from any proportions of different phenotypes, the final proportions always converge to the same constants. For example, breast cancer has three cell phenotypes: stem-like, basal, and luminal. Starting from any one phenotype, after cultivating for some time, the final proportions are always stem-like $2\%$, basal $97\%$, luminal $1\%$ \cite{gupta2011stochastic}. Therefore, one can perform corresponding experiments to detect the existence of such phenomena, and determine whether the cancer cell type in the experiment has conversions from the non-stem state to the stem state.

In the following, we will review different models for cancer cell population dynamics, and corresponding explanations for the phenotypic equilibrium phenomena. We consider $n$ phenotypes in the cancer cell population: $Y_{1},Y_{2},\cdots,Y_{n}$. Define $\vec{X}(t)=[X_1(t),X_2(t),\cdots, X_n(t)]$ to be the population at time $t$, where $X_i(t)$ is the population of phenotype $Y_i$. $P_i(t)=X_i(t)/[\sum_{i=1}^n X_i(t)]$ is the proportion of phenotype $Y_i$, which is not defined if $\sum_{i=1}^n X_i(t)=0$. Define $\vec{P}(t)=[P_1(t),P_2(t),\cdots,P_n(t)]$.

\subsubsection{Markov chain model} The first explanation for the phenotypic equilibrium phenomena is based on a Markov chain model \cite{gupta2011stochastic}. In this model, $\vec{P}(t)$ satisfies 
\[\mathrm{d}\vec{P}(t)/\mathrm{d}t=\vec{P}(t)\mathbf{Q},\]
where $\mathbf{Q}$ is the transition rate matrix, satisfying $\vec{1}\mathbf{Q}'=\vec{0}$. This is the evolution equation for a Markov chain. From classical Markov chain theory, we have the following \cite{norris1998markov}: If $\mathbf{Q}$ is irreducible, then starting from any initial probability distribution $\vec{P}(0)$, $\vec{P}(t)$ always converges to the same vector $\vec{v}$, which satisfies $\vec{v}\mathbf{Q}=\vec{v}$. This explains the phenotypic equilibrium phenomena in the Markov chain model. However, this model has been criticized as unrealistic \cite{jiang2017phenotypic}, since cells can grow, and different phenotypes might have different growth rates.

\subsubsection{ODE model} A classic model for population dynamics is the linear ODE model. We assume that the population (not the proportion) satisfies 
\[\mathrm{d}\vec{X}(t)/\mathrm{d}t=\vec{X}(t)\mathbf{A},\]
where $\mathbf{A}=\{a_{i,j}\}$ for $i,j=1,\ldots,n$, and $a_{i,j}\ge 0$ for $i\ne j$ \cite{zhou2014multi}. To study the limiting behavior of $\vec{X}(t)$, we first need a result in matrix theory \cite{karlin2014first}:
\begin{theorem}
	$\mathbf{A}$ has a real eigenvalue $\lambda_1$, such that for any eigenvalue $\mu\ne \lambda_1$, $\Re(\mu)<\lambda_1$. $\lambda_1$ has a left eigenvector $\vec{u}=(u_1,\ldots ,u_n)$ that satisfies $u_i\ge 0$ and $\sum_{i=1}^n u_i=1$.
	\label{thm2}
\end{theorem}
Now we can present the convergence result in the ODE model \cite{zhou2014multi}:
\begin{theorem}
	Assume that $\lambda_1$ is a simple eigenvalue of $\mathbf{A}$. Starting from any initial value $\vec{X}(0)$ except for a zero-measure set, we have $\lim_{t \to \infty}\vec{P}(t)\to \vec{u}$. 
\end{theorem}
Therefore, the proportion $\vec{P}(t)$ will converge to the unique attracting fixed point $\vec{u}$. This explains the phenotypic equilibrium phenomena in the ODE model. Nevertheless, on single cell level, the cell growth is essentially stochastic. The ODE model is just a mass action approximation.

\subsubsection{Branching process model} A more realistic model for cell population dynamics is the multi-type branching process model. In this model, different cells at the same time point are independent. For each cell of the $Y_i$ type, the behavior is described by the following: 
\[Y_{i}\stackrel{\alpha_i}{\to}d_{i1}Y_{1}+d_{i2}Y_{2}+\cdots{}+d_{in}Y_{n}.\]
Therefore, a $Y_i$ cell lives an exponential time with expectation $1/\alpha_i$ and turns into $d_{i1}$ $Y_{1}$ cells, $d_{i2}$ $Y_{2}$ cells, $\cdots$, $d_{in}$ $Y_{n}$ cells, where $d_{i1},d_{i2},\cdots{},d_{in}$ are random variables in $\mathbb{N}$. The random coefficients $d_{i1},d_{i2},$ $\cdots{},d_{in}$ may not be independent, but they are independent with the exponential waiting time. For example, a symmetric division $Y_1\to 2Y_1$ means $(d_{11},d_{12},\cdots,d_{1n})=(2,0,0,\cdots,0)$. A transition $Y_1\to Y_2$ means $(d_{11},d_{12},\cdots,d_{1n})=(0,1,0,\cdots,0)$. A death $Y_1\to \emptyset$ means $(d_{11},d_{12},\cdots,d_{1n})=(0,0,0,\cdots,0)$. Define \[a_{i,i}=\alpha_i(\mathbb{E}d_{ii}-1)\ge-\alpha_i,\ \ \ a_{i,j}=\alpha_j\mathbb{E}d_{ij} \text{ for } i\ne j.\] 
Then the matrix $\mathbf{A}=\{a_{i,j}\}$ corresponds to the coefficient matrix in the ODE model. We have the same $\lambda_1$ and $\vec{u}$ as in Theorem~\ref{thm2}.

In the branching process model, due to randomness, we need to assume that no phenotype $Y_i$ has $X_i(t)=0$ for any $t>T$. This condition is called non-extinction. We can prove the convergence of $\vec{P}(t)$, namely a law of large numbers, conditioned on non-extinction. In a monograph on branching process in 1972, a law of large numbers is proven for certain $\mathbf{A}$ \cite{athreya2004branching}:
\begin{theorem}
	Assume that $\lambda_1$ is positive, and it is a simple eigenvalue of $\mathbf{A}$. Assume $\mathbf{A}$ is irreducible. Conditioned on non-extinction, $\vec{P}(t) \to \vec{u}$ almost surely.
\end{theorem}
This law of large numbers can be used to explain the phenotypic equilibrium phenomena. Nevertheless, the irreducibility of $\mathbf{A}$ means that any two phenotypes can convert between each other (might through some other phenotypes). This might not hold for some cancer types. The irreducibility assumption is partially lifted in a 2004 paper \cite{janson2004functional} (only assume that the phenotype with the highest growth rate can convert to any other phenotype), and fully lifted in a 2017 paper \cite{jiang2017phenotypic}:
\begin{theorem}
	Assume that $\lambda_1$ is positive, and it is a simple eigenvalue of $\mathbf{A}$. Conditioned on non-extinction, $\vec{P}(t) \to \vec{u}$ almost surely.
\end{theorem}
This explains the phenotypic equilibrium phenomena in the branching process model.

In the branching process model, the waiting time for cell division is exponential, which is not realistic. To overcome this flaw, one can generalize the law of large numbers to non-Markovian branching processes. The trick is to add auxiliary phenotypes, so that the overall waiting time is close to the real distribution. Nevertheless, this trick is not rigorous, and general cases are still open. This illustrates that the need in cancer research might inspire new mathematical results.

\subsection{Inferring the existence of multiple phenotypes}
\label{s2.2}
\begin{figure}[ht]
	\centering
	\includegraphics[width=0.9\linewidth]{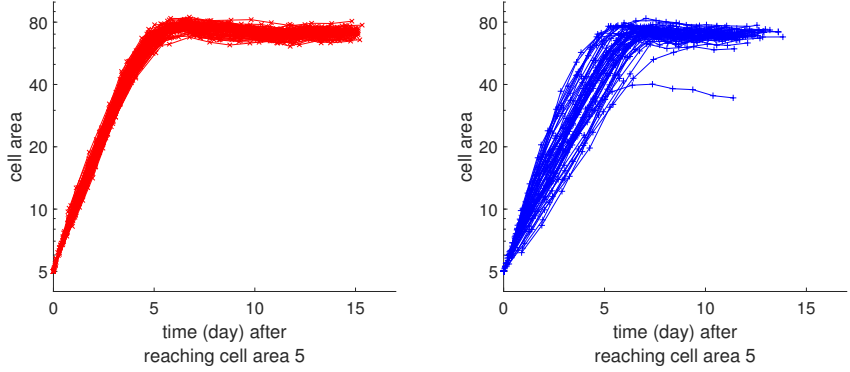}  
	\caption{Growth curves of cell populations starting from the time of reaching 5 area units, with a logarithm scale for the $y$-axis. The $x$-axis is the time from reaching 5 area units. Red (left) and blue (right) curves correspond to 10 or 1 initial cell(s). Although starting from the same population level, patterns are different for distinct initial cell numbers. The $N_0=1$-cell group has higher diversity and lower average population level. This figure corresponds to the contents of Subsection~\ref{s2.2}.
	}
	\label{f1}
\end{figure}

\begin{figure}[ht]
	\centering
	\includegraphics[width=0.9\linewidth]{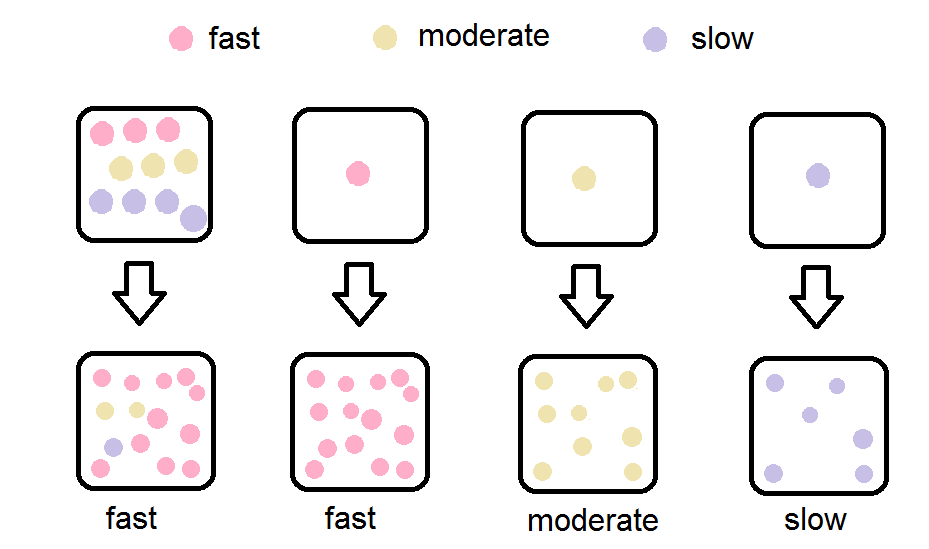}  
	\caption{Schematic illustration of the qualitative argument: Three cell types and growth patterns (three colors) with different seeding numbers. One $N_0=10$-cell well will have at least one fast type cell with high probability, which will dominate the population. One $N_0=1$-cell well can only have one cell type, thus in the microculture ensemble of replicate wells, three possible growth patterns for wells can be observed. This figure corresponds to the contents of Subsection~\ref{s2.2}.}
	\label{f2}
\end{figure}

The experiments for phenotypic equilibrium need to measure the cell states. In fact, even without knowing the cell state, but just counting cell numbers, we can infer the existence of multiple phenotypes. In some experiments, leukemia cells were cultivated. Some populations started from 10 cells, while other populations started from 1 cell. The cell number of each population was measured everyday. By comparing cell population levels, we can find that after reaching the same population level, the number of initial cells still affects cell behavior. Specifically, when reaching cell area 5 (approximately 2500 cells), the growth rates of populations growing from 10 initial cells are higher than those of populations from 1 initial cell. See Fig.~\ref{f1} for translated growth curves. This difference indicates the existence of multiple phenotypes with different growth rates: starting from 10 initial cells, it is very likely that at least one cell is of the fast type, which will dominate the population; starting from 1 initial cell, the population might only have the slow type, which corresponds to a lower growth rate \cite{wang2023multiple}. See Fig.~\ref{f2} for a schematic illustration. Certainly, this explanation needs to be verified by further experiments that measure the phenotypes, especially with gene sequencing. Without such data, one needs to run a parameter scan to confirm that the model is robust.

\subsection{Cancer cell under treatment}
Various drugs can be used to kill cancer cells. Besides traditional chemical drugs, aptamers \cite{wang2022discrete} and T cells \cite{dessalles2021naive} can also be used. Therefore, the population dynamics of cancer cells under treatment is also worth studying.

The behavior of cancer under treatment can be complicated. The tumor cell regrowth after treatment is generally regarded as the result of
a Darwinian somatic evolution: Given sufficient genetic variability in a sufficiently large initial cell population, it is very likely that the population contains cells with genomic mutations that lead to drug-resistance. A single cell with such a mutation will survive the treatment and clonally expand, thus driving the tumor regrowth under treatment. If this simple Darwinian model holds, then the recurrence time of cancer should be positively related to the drug dose. The reason is that at a higher drug dose, the number of survived cells is smaller, thus taking a longer time to recover. However, in some experiments, it is observed that the treatment effect does not always become better when the drug dose increases. The treatment effect gradually saturates. If the drug dose is sufficiently high, the cancer cell population recovers even faster. Due to such phenomena, we have to introduce more complicated mechanisms. One possible approach is to assume that at higher drug doses, the killing rate does not further increase, but the drug-induced transition rate from the sensitive state to the resistant state still increases \cite{angelini2022model}. Therefore, when we increase the drug dose, the treatment effect first increases, since the killing rate increases. Then the killing rate is almost saturated, and the treatment effect does not further increase. Finally, the killing rate stops at its maximum, and the transition rate still increases, which decreases the treatment effect.

Besides the phenomena that higher drug dose leads to worse treatment effects, decreasing the drug dose might lead to overcompensation, which means that the population recovers to a higher level than the initial population \cite{xia2023age}. Therefore, the treatment of cancer is a highly complicated decision process. It is necessary to introduce reinforcement learning (especially with stochasticity) into this field \cite{tseng2017deep,wang2021measuring,zhao2009reinforcement,wang2023online,zhao2011reinforcement}.

\section{Gene regulation}
\label{sec3}
Coding genes are transcribed to mRNAs and translated to proteins. Genes (through their corresponding proteins) can regulate the expressions of other genes. The origin of cancer is certain genetic mutations (e.g., p53) \cite{hafner2019multiple}. Such mutations change the gene expression patterns through gene regulation \cite{statello2021gene}, and the cell landscape also transforms \cite{li2019landscape}. Therefore, the study of gene regulation is important in understanding and curing cancer \cite{ballestar2008epigenetic}.  

\subsection{Framework for gene regulation inference}
\label{s3.1}
Genes and their regulatory relations form gene regulatory networks (GRNs). Since the gene expression and regulation are mostly confined within living cells, it is almost impossible to determine gene regulation with direct biochemical methods. Nevertheless, there have been many methods to infer the GRN structure, namely whether one gene regulates another, from experimental data \cite{bansal2006inference,bocci2022splicejac,hurley2012gene}. Such inference methods use models based on different assumptions, and need different types of data. To unify different approaches for this problem, there is a framework for data classification, which has four major dimensions: (1) whether measuring gene expression levels, or the levels of certain phenotypes that are determined by such genes; (2) whether the measurement is for a single cell (stochastic) or for a large population of cells (deterministic); (3) whether the measurement is for one time point or multiple time points; (4) whether one can add interventions on certain genes. If the measurement is on single-cell level and at multiple time points, depending on whether the same cell can be measured multiple times, there is an extra dimension of whether we have the joint probability distribution or just marginal probability distributions. In total, there are 20 different data types \cite{wang2022inference}. Different data types need different inference methods, and some data types (especially non-interventional) are not useful in inferring gene regulation. Due to the large amount of related data, machine learning is a promising future direction for inferring gene regulatory relations \cite{huynh2010inferring}. See Table~\ref{Tab} for the classification of data types and corresponding methods that can be applied to infer gene regulations.

\begin{table}[]
	\begin{tabular}{|l|l|l|l|l|l|}
		\hline
		\multicolumn{2}{|l|}{\multirow{2}{*}{}}                                                                                             & \multicolumn{2}{l|}{One-Time}                                                                                                                                                                   & \multicolumn{2}{l|}{Time Series}                                                                                                                                                                                                                                                             \\ \cline{3-6} 
		\multicolumn{2}{|l|}{}                                                                                                              & \begin{tabular}[c]{@{}l@{}}Non-\\ Intervention\end{tabular}                                & Intervention                                                                                       & \begin{tabular}[c]{@{}l@{}}Non-\\ Intervention\end{tabular}                                                                   & Intervention                                                                                                                                                 \\ \hline
		\multirow{2}{*}{\begin{tabular}[c]{@{}l@{}}Gene\\ Expression\end{tabular}} & \begin{tabular}[c]{@{}l@{}}Single-\\ Cell\end{tabular} & \begin{tabular}[c]{@{}l@{}}Scenario 1\\ \\ MF+DAG: \\ partial\\ \end{tabular} & \begin{tabular}[c]{@{}l@{}}Scenario 2\\ \\ PB: full\\ DAG: partial\\ MF+DAG: \\ full\\ \end{tabular} & \begin{tabular}[c]{@{}l@{}}Scenario 3 a/b\\ \\ 3a Joint:\\ UC: full\\  3b Marginal:\\ MF+DAG: \\ partial\\ \end{tabular} & \begin{tabular}[c]{@{}l@{}}Scenario 4 a/b\\ \\ 4a Joint:\\ UC: full\\  4b Marginal:\\ LS: full\\ PB: full\\ DAG: partial\\ MF+DAG: full\\ \end{tabular} \\ \cline{2-6} 
		& Bulk                                                   & \begin{tabular}[c]{@{}l@{}}Scenario 5\\ \\ No\end{tabular}                                 & \begin{tabular}[c]{@{}l@{}}Scenario 6\\ \\ PB: full\\ DAG: partial\end{tabular}                    & \begin{tabular}[c]{@{}l@{}}Scenario 7\\ \\ No\end{tabular}                                                                    & \begin{tabular}[c]{@{}l@{}}Scenario 8\\ \\ LS: full\\ PB: full\\ DAG: partial\end{tabular}                                                                   \\ \hline
		\multirow{2}{*}{Phenotype}                                                 & \begin{tabular}[c]{@{}l@{}}Single-\\ Cell\end{tabular} & \begin{tabular}[c]{@{}l@{}}Scenario 9\\ \\ No\end{tabular}                                 & \begin{tabular}[c]{@{}l@{}}Scenario 10\\ \\ PB: partial\end{tabular}                               & \begin{tabular}[c]{@{}l@{}}Scenario 11 a/b\\ \\ No\end{tabular}                                                                   & \begin{tabular}[c]{@{}l@{}}Scenario 12 a/b\\ \\ PB: partial\\ LS+DAG: \\ partial*\\ PB+LS+DAG: \\ partial*\end{tabular}                                                                    \\ \cline{2-6} 
		& Bulk                                                   & \begin{tabular}[c]{@{}l@{}}Scenario 13\\ \\ No\end{tabular}                                & \begin{tabular}[c]{@{}l@{}}Scenario 14\\ \\ PB: partial\end{tabular}                               & \begin{tabular}[c]{@{}l@{}}Scenario 15\\ \\ No\end{tabular}                                                                   & \begin{tabular}[c]{@{}l@{}}Scenario 16\\ \\ PB: partial\\ LS+DAG: \\ partial*\\ PB+LS+DAG: \\ partial*\end{tabular}                                                                    \\ \hline
	\end{tabular}
	\caption{GRN structure inference with different data types: under what assumptions, what structures can be inferred. There are 16 scenarios classified by the following dimensions of data types: Gene Expression vs. Phenotype; Single-Cell vs. Bulk; One-Time vs. Time Series; Non-Interventional vs. Interventional. In Scenarios 3/4/11/12, there is an extra dimension of Joint vs. Marginal. There are different assumptions: PB: path blocking; DAG: directed acyclic graph; MF: Markov and faithful; LS: linear system; UC: unconditional. Full/partial/no means all/some/no GRN structures can be inferred. For example, ``MF+DAG: partial'' means under MF assumption and DAG assumption, GRN structure can be partially inferred. The asterisk * in Scenarios 12/16 means for some identified regulatory relations, we cannot determine whether they are activation or inhibition \cite{wang2022inference}. This table corresponds to the contents of Subsection~\ref{s3.1}.}
	\label{Tab}
\end{table}

\subsection{Inference methods for mutual regulation}
We briefly introduce some inference methods for mutual regulation. Each method may apply in multiple scenarios. Different methods need different assumptions. DAG assumption means that the GRN is a directed graph that has no
directed cycles. MF assumption means that the joint probability distribution of gene expressions provide exactly all and only the conditional independence relations implied by the GRN. LS assumption means that the gene expression satisfies a linear ODE system. PB assumption means that if all paths between two genes in a GRN have been blocked by intervened genes, then adding intervention on one gene cannot further affect the other gene. 

\subsubsection{Causal DAG method}
This method needs the MF and DAG assumptions, and applies to Scenarios 1 and 3b \cite{glymour2016causal}. In this case, the GRN is a causal DAG, and we can use the conditional independence relations to determine the corresponding causal DAG. However, some different causal DAGs (all sharing the same edges, but with different directions) produce the same conditional independence relations. Therefore, we can only partially determine the GRN structure.

\subsubsection{Path blocking relation method in gene expression}
This method needs the PB assumption, and applies to Scenarios 2, 4b, 6, 8 \cite{wang2022inference}. From such data types, we know whether some genes fully block all paths between two genes. There is an edge $G_1\to G_2$ if and only if all other genes cannot block the path from $G_1$ to $G_2$. All edges can be inferred in this case.

\subsubsection{Ancestor-descendant relation method}
This method needs the DAG assumption, and applies to Scenarios 2, 4b, 6, 8 \cite{wang2022inference}. From such data types, we can use interventions to determine whether one gene $G_1$ has a directed path to another gene $G_2$ in the GRN (ancestor-descendant relations). We can determine an equivalent class that share the same ancestor-descendant relations by the following theorem:
\begin{theorem}
	The following procedure describes how to determine certain edges with ancestor-descendant relations. (1) If $G_j$ is not a descendant of $G_i$, then we can determine that the edge $G_i\to G_j$ does not exist. (2) If $G_j$ is a descendant of $G_i$, and $G_i$ has another descendant $G_k$, which is an ancestor of $G_j$, then we cannot determine the existence of the edge $G_i\to G_j$, since we can find two GRNs with the same ancestor-descendant relations, one with $G_i\to G_j$, and one without $G_i\to G_j$. (3) If $G_j$ is a descendant of $G_i$, and $G_i$ does not have another descendant $G_k$, which is an ancestor of $G_j$, then we can determine that the edge $G_i\to G_j$ exists. 
\end{theorem}
Therefore, we can partially determine the GRN structure. About the number of inferred edges by this method, we have a lower bound \cite{wang2022inference}: 
\begin{theorem}
	If the GRN is a connected DAG with $n$ genes, then we can use ancestor-descendant relations to identify at least $n-1$ edges. 
\end{theorem}

\subsubsection{Ancestor-descendant relation and causal DAG method}
This method needs the MF and DAG assumptions, and applies to Scenarios 2 and 4b \cite{wang2022inference}. By the causal DAG method, we can determine all edges in the GRN, but the directions of some edges are unknown. Using the ancestor-descendant relation, we can determine such directions. Therefore, the GRN can be fully determined.

\subsubsection{Causal time series method}
This method does not need any assumption, and applies to Scenarios 3a and 4a \cite{glymour2016causal}. This is basically the same as the causal DAG method, but the time series data naturally determines the direction of a causal relation. Therefore, we can use conditional independence relations to fully determine the GRN structure.

\subsubsection{Linear ODE method}
This method needs the LS assumption, and applies to Scenarios 4b and 8 \cite{pollicott2012extracting}. We have a linear system for the gene expression levels, each with the form 
\[\mathrm{d}X_i/\mathrm{d}t=\sum_{j=1}^n a_{ji}X_j.\]
We can measure $X_i(t)$ at different time points under different interventions, and calculate $\mathrm{d}X_i/\mathrm{d}t$. Then we obtain a linear algebraic equation system for $a_{ji}$. We can solve this and determine the GRN structure.

\subsubsection{Path blocking relation method in phenotype}
This method needs the PB assumption, and applies to Scenarios 10, 12a/b, 14, 16 \cite{wang2022inference}. From such data types, we know whether some genes fully block all paths from a gene $G_i$ to a phenotype $P$. If a subset $\mathcal{S}$ of $\{G_1,\ldots,G_n\}\backslash\{G_i\}$ blocks $G_i$ to $P$, but any proper subset of $\mathcal{S}$ cannot block $G_i$ to $P$, then $\mathcal{S}$ is called a minimal blocking set. If a blocking set $\mathcal{S}$ is not minimal, then $\mathcal{S}$ contains a blocking subset that is minimal. Define $\beta(G_i)$ to be all minimal subsets that block $G_i$ to $P$. The following theorem describes what edges can or cannot be inferred from the path blocking relations \cite{wang2022inference}.
\begin{theorem}
	The following procedure describes how to determine certain edges with path blocking relations for the phenotype. (1) There is an edge $G_i\to P$ if and only if $\beta(G_i)=\emptyset$. (2) If there exists $\mathcal{S}\in\beta(G_i)$, so that $G_j\notin \mathcal{S}$, and $\mathcal{S}$ cannot block $G_j$ to $P$, then there is no edge $G_i\to G_j$. (3) If (2) is not satisfied, but there exists $\mathcal{S}\in\beta(G_i)$, so that $G_j\in \mathcal{S}$, then there is an edge $G_i\to G_j$. (4) If $\beta(G_i)=\emptyset$, or for any $\mathcal{S}\in\beta(G_i)$, we have $G_j\notin \mathcal{S}$, and $\mathcal{S}$ blocks $G_j$ to $P$, then we cannot determine whether $G_i\to G_j$ exists, since we can find two GRNs with the same path blocking relations, one with $G_i\to G_j$, and one without $G_i\to G_j$.
\end{theorem}
Using the above procedure, we can partially determine the GRN structure. About the number of inferred edges by this method, we have a lower bound \cite{wang2022inference}: 
\begin{theorem}
	If the GRN with $n$ genes has a directed path from each gene $G_i$ to the phenotype $P$, then we can use path blocking relations for the phenotype to identify at least $n$ edges. 
\end{theorem}

\subsection{Inference methods for autoregulation}
Some genes can regulate (activate or inhibit) their own expressions, which is called autoregulation. Certain types of data can be used to infer the existence of gene autoregulation, such as time series data \cite{sanchez2018bayesian,xing2005causal,veerman2022parameter} or interventional data for single-cell gene expression \cite{jia2018relaxation}. There also exists an inference method that only requires one-time non-interventional single-cell gene expression data \cite{wang2022ar}. Although different inference methods are based on different types of models, and require different data types, the basic idea is the same: build two models, one with autoregulation, and one without autoregulation; then try to find different behaviors between these two models. This idea can be generalized to other problems of detecting the existence of certain mechanisms. If these two models are indistinguishable, it is not a negative result either: one proves that this mechanism is not detectable in this setting \cite{cao2018linear}! Here we introduce a simple Markov chain model for gene expression and present an inference method for autoregulation.

Consider $m$ genes $G_1,\ldots,G_m$ for a single cell. Denote their expression levels by random variables $X_1,\ldots,X_m$. Each state of this Markov chain, $(X_1=n_1,\ldots,X_i=n_i,\ldots,X_m=n_m)$, can be abbreviated as $\boldsymbol{n}=(n_1,\ldots,n_i,\ldots,n_m)$. For gene $G_i$, the transition rate of $n_i-1\to n_i$ is $f_i(\boldsymbol{n})$, and the transition rate of $n_i\to n_i-1$ is $g_i(\boldsymbol{n})n_i$. Transitions with more than one step are not allowed. The master equation of this process is 
\begin{equation*}
	\label{me}
	\begin{split}
		\frac{\mathrm{d}\mathbb{P}(\boldsymbol{n})}{\mathrm{d}t}=&\sum_i\mathbb{P}(n_1,\ldots,n_i+1,\ldots,n_m)g_i(n_1,\ldots,n_i+1,\ldots,n_m)(n_i+1)\\
		+&\sum_i\mathbb{P}(n_1,\ldots,n_i-1,\ldots,n_m)f_i(\boldsymbol{n})\\
		-&\mathbb{P}(\boldsymbol{n})\sum_i[f_i(n_1,\ldots,n_i+1,\ldots,n_m)+g_i(\boldsymbol{n})n_i].
	\end{split}		
\end{equation*}
Define $\boldsymbol{n}_{\bar{i}}=(n_1,\ldots,n_{i-1},n_{i+1},\ldots,n_m)$. Define $h_i(\boldsymbol{n})=f_i(\boldsymbol{n})/g_i(\boldsymbol{n})$ to be the relative growth rate of gene $V_i$. Autoregulation means for some fixed $\boldsymbol{n}_{\bar{i}}$,  $h_i(\boldsymbol{n})$ is (locally) increasing/decreasing with $n_i$, thus $f_i(\boldsymbol{n})$ increases/decreases and/or $g_i(\boldsymbol{n})$ decreases/increases with $n_i$. In this model, we have the following result \cite{wang2022ar}:
\begin{theorem}
	In the above model, assume the GRN has no directed cycle that contains gene $G_k$. Assume $g_k(\cdot)$ is a constant for all $\boldsymbol{n}$. If $G_k$ has no autoregulation, meaning that $h_k(\cdot)$ and $f_k(\cdot)$ do not depend on $n_k$, then the expression level of $G_k$ satisfies $\mathrm{var}(X_k)/\mathbb{E}(X_k)\ge 1$. Therefore, $G_k$ has $\mathrm{var}(X_k)/\mathbb{E}(X_k)< 1$ means $G_k$ has autoregulation.
\end{theorem}
Nevertheless, this method fails to determine whether autoregulation exists when we have $\mathrm{var}(X_k)/\mathbb{E}(X_k)\ge 1$ for one gene. It is hypothesized that the requirement for $g_k(\cdot)$ can be dropped:
\begin{conjecture}
	In the above model, assume the GRN has no directed cycle that contains gene $G_k$. If $h_k(\cdot)$ and $f_k(\cdot)$ do not depend on $n_k$, then the expression level of $G_k$ satisfies $\mathrm{var}(X_k)/\mathbb{E}(X_k)\ge 1$. 
\end{conjecture}

\subsection{Other topics in gene regulation}
If we cannot add intervention, and cannot measure the same cell(s) multiple times, then data from different time points are essentially the same, and we cannot obtain information of the dynamics. In this case, sometimes it is impossible to build the causal relationship in gene regulation (including autoregulation) \cite{wang2020causal}.

With the inferred knowledge of gene regulations, one can explain some phenomena in cancer biology. For example, myeloproliferative neoplasm needs two mutations, JAK2 and TET2. The order of the appearance of these two mutations can affect clinical features and responses to therapy \cite{ortmann2015effect,levine2019roles,kent2017order}. Possible explanations include threshold phenomena and hidden variables.

Gene regulation has many complicated mechanisms. Certain genes can translocate in gene sequences, which are called transposons. Genes at different locations can have different effects in gene regulation \cite{medstrand2005impact}. Therefore, it is also important to design methods to determine transposons in gene sequences \cite{wang2023longest,kang2014flexibility,gardner2017mobile,chen2019ervcaller,yu2021benchmark}. The general idea is to find the longest common subsequence, whose complement is all the transposons. Determining the longest common subsequence is a classic problem in computer science.

\section{Developmental biology}
\label{sec4}
Developmental biology studies how a zygote differentiates into normal tissues, while cancer biology studies how normal cells become malfunctional. Seemingly in opposite directions, these two subjects are interconnected. The microenvironment of embryo can suppress the emergence of cancer \cite{biava2002cancer}. In fact, certain cancer cells transplanted to an embryo will differentiate into normal tissues \cite{postovit2008human}. The development of an embryo is extremely complicated, so that a cumbersome model is still an oversimplification \cite{wang2020model}.

\subsection{Inference on transplantation experiments}
\label{s4.1}
The transplantation experiments of cancer cells and embryos indicate a possible direction of cancer treatment. However, such transplantation experiments are expensive and difficult to perform. Therefore, inference methods that determine unknown experimental results from known results are necessary \cite{wang2021inference}. The idea is inspired by Ising model in statistical physics. Each transplantation experiment is determined by the donor tissue and the host tissue. We can construct the similarities between tissues. Then transplantation experiments can be compared by the similarities between their donor tissues and host tissues. We can assume that similar experiments are more likely to have similar results. Given some transplantation experiments, where some results are known, and some results are unknown, we can take a guess for those unknown results. Then we feed all results to a penalty function. Assume the transplantation experiments are $X_1,\ldots,X_n$; their results are $x_1,\ldots,x_n$; the similarity function between two transplantation experiments is $F(X_i,X_j)$; the similarity function between results is $f(x_i,x_j)$. Then the penalty function should be 
\[P(x_1,\ldots,x_n)=-\sum_{i=1}^n \sum_{j=1}^n F(X_i,X_j)f(x_i,x_j).\]
Therefore, for each result configuration $(x_1,\ldots,x_n)$, we can calculate its penalty. The penalty can be mapped to its probability:
\[\mathbb{P}(x_1,\ldots,x_n)=\frac{e^{-\beta P(x_1,\ldots,x_n)}}{\sum_{x_1',\ldots,x_n'}e^{-\beta P(x_1',\ldots,x_n')}},\]
where $\beta$ is a parameter that should be determined. We can output the result configuration with the largest probability, or take expectation for all configurations to output the probability distribution for each experiment. In some situations, known results are not deterministic, but stochastic. For example, assume each experiment has two possible results: $N$ (normal) and $A$ (abnormal). Assume one experiment has result $\mathbb{P}(N)=1/3$, and the other experiment has result $\mathbb{P}(N)=2/3$. We can assume those two experiments are independent, and sample deterministic results. Then $\mathbb{P}(N,N)=2/9$, $\mathbb{P}(N,A)=1/9$, $\mathbb{P}(A,N)=4/9$, $\mathbb{P}(A,A)=2/9$. For each possible configuration, apply the inference method, and then take expectations for all configurations.

The above inference method can be applied in the opposite direction. Assume we want to know the results of some transplantation experiments. We only need to perform some experiments, and use the inference method to determine other results. Now the question is to design experiments, so that all the results can be determined, while the cost is minimized. After transforming the similarities between experiments into the similarity chart, the experiment design problem becomes a coloring problem. See Table~\ref{co} for the optimal coloring configurations in two-dimensional lattice, where each not colored unit (not conducted experiment) is neighboring to $4/2/1$ colored units (conducted experiment). Since the actual transplantation experiment should be described by at least four factors (donor type, donor stage, host type, host stage), the coloring problem should be generalized to any $n$-dimensional lattice.

\begin{table}[]
	\begin{tabular}{lllllllll}
		&                         &                        &                        &                        & \multicolumn{2}{l}{Donor}                       &                        &                        \\
		&                         & $T_1$                     & $T_2$                     & $T_3$                     & $T_4$                     & $T_5$                     & $T_6$                    & $T_7$                     \\ \cline{3-9} 
		& \multicolumn{1}{l|}{$T_1$} & \multicolumn{1}{l|}{\cellcolor{black}} & \multicolumn{1}{l|}{}  & \multicolumn{1}{l|}{\cellcolor{black}} & \multicolumn{1}{l|}{}  & \multicolumn{1}{l|}{\cellcolor{black}} & \multicolumn{1}{l|}{}  & \multicolumn{1}{l|}{\cellcolor{black}} \\ \cline{3-9} 
		& \multicolumn{1}{l|}{$T_2$} & \multicolumn{1}{l|}{}  & \multicolumn{1}{l|}{\cellcolor{black}} & \multicolumn{1}{l|}{}  & \multicolumn{1}{l|}{\cellcolor{black}} & \multicolumn{1}{l|}{}  & \multicolumn{1}{l|}{\cellcolor{black}} & \multicolumn{1}{l|}{}  \\ \cline{3-9} 
		& \multicolumn{1}{l|}{$T_3$} & \multicolumn{1}{l|}{\cellcolor{black}} & \multicolumn{1}{l|}{}  & \multicolumn{1}{l|}{\cellcolor{black}} & \multicolumn{1}{l|}{}  & \multicolumn{1}{l|}{\cellcolor{black}} & \multicolumn{1}{l|}{}  & \multicolumn{1}{l|}{\cellcolor{black}} \\ \cline{3-9} 
		Host & \multicolumn{1}{l|}{$T_4$} & \multicolumn{1}{l|}{}  & \multicolumn{1}{l|}{\cellcolor{black}} & \multicolumn{1}{l|}{}  & \multicolumn{1}{l|}{\cellcolor{black}} & \multicolumn{1}{l|}{}  & \multicolumn{1}{l|}{\cellcolor{black}} & \multicolumn{1}{l|}{}  \\ \cline{3-9} 
		& \multicolumn{1}{l|}{$T_5$} & \multicolumn{1}{l|}{\cellcolor{black}} & \multicolumn{1}{l|}{}  & \multicolumn{1}{l|}{\cellcolor{black}} & \multicolumn{1}{l|}{}  & \multicolumn{1}{l|}{\cellcolor{black}} & \multicolumn{1}{l|}{}  & \multicolumn{1}{l|}{\cellcolor{black}} \\ \cline{3-9} 
		& \multicolumn{1}{l|}{$T_6$} & \multicolumn{1}{l|}{}  & \multicolumn{1}{l|}{\cellcolor{black}} & \multicolumn{1}{l|}{}  & \multicolumn{1}{l|}{\cellcolor{black}} & \multicolumn{1}{l|}{}  & \multicolumn{1}{l|}{\cellcolor{black}} & \multicolumn{1}{l|}{}  \\ \cline{3-9} 
		& \multicolumn{1}{l|}{$T_7$} & \multicolumn{1}{l|}{\cellcolor{black}} & \multicolumn{1}{l|}{}  & \multicolumn{1}{l|}{\cellcolor{black}} & \multicolumn{1}{l|}{}  & \multicolumn{1}{l|}{\cellcolor{black}} & \multicolumn{1}{l|}{}  & \multicolumn{1}{l|}{\cellcolor{black}} \\ \cline{3-9} 
		&                         &                        &                        &                        & \multicolumn{2}{l}{}                            &                        &                        \\
		&                         & T1                     & T2                     & T3                     & T4                     & T5                     & T6                     & T7                     \\ \cline{3-9} 
		& \multicolumn{1}{l|}{T1} & \multicolumn{1}{l|}{}  & \multicolumn{1}{l|}{\cellcolor{black}}& \multicolumn{1}{l|}{}  & \multicolumn{1}{l|}{}  & \multicolumn{1}{l|}{\cellcolor{black}}& \multicolumn{1}{l|}{}  & \multicolumn{1}{l|}{}  \\ \cline{3-9} 
		& \multicolumn{1}{l|}{T2} & \multicolumn{1}{l|}{\cellcolor{black}}& \multicolumn{1}{l|}{}  & \multicolumn{1}{l|}{}  & \multicolumn{1}{l|}{\cellcolor{black}}& \multicolumn{1}{l|}{}  & \multicolumn{1}{l|}{}  & \multicolumn{1}{l|}{\cellcolor{black}}\\ \cline{3-9} 
		& \multicolumn{1}{l|}{T3} & \multicolumn{1}{l|}{}  & \multicolumn{1}{l|}{}  & \multicolumn{1}{l|}{\cellcolor{black}}& \multicolumn{1}{l|}{}  & \multicolumn{1}{l|}{}  & \multicolumn{1}{l|}{\cellcolor{black}}& \multicolumn{1}{l|}{}  \\ \cline{3-9} 
		Host & \multicolumn{1}{l|}{T4} & \multicolumn{1}{l|}{}  & \multicolumn{1}{l|}{\cellcolor{black}}& \multicolumn{1}{l|}{}  & \multicolumn{1}{l|}{}  & \multicolumn{1}{l|}{\cellcolor{black}}& \multicolumn{1}{l|}{}  & \multicolumn{1}{l|}{}  \\ \cline{3-9} 
		& \multicolumn{1}{l|}{T5} & \multicolumn{1}{l|}{\cellcolor{black}}& \multicolumn{1}{l|}{}  & \multicolumn{1}{l|}{}  & \multicolumn{1}{l|}{\cellcolor{black}}& \multicolumn{1}{l|}{}  & \multicolumn{1}{l|}{}  & \multicolumn{1}{l|}{\cellcolor{black}}\\ \cline{3-9} 
		& \multicolumn{1}{l|}{T6} & \multicolumn{1}{l|}{}  & \multicolumn{1}{l|}{}  & \multicolumn{1}{l|}{\cellcolor{black}}& \multicolumn{1}{l|}{}  & \multicolumn{1}{l|}{}  & \multicolumn{1}{l|}{\cellcolor{black}}& \multicolumn{1}{l|}{}  \\ \cline{3-9} 
		& \multicolumn{1}{l|}{T7} & \multicolumn{1}{l|}{}  & \multicolumn{1}{l|}{\cellcolor{black}}& \multicolumn{1}{l|}{}  & \multicolumn{1}{l|}{}  & \multicolumn{1}{l|}{\cellcolor{black}}& \multicolumn{1}{l|}{}  & \multicolumn{1}{l|}{}  \\ \cline{3-9} 
		&                         &                        &                        &                        & \multicolumn{2}{l}{}                            &                        &                        \\
		&                         & T1                     & T2                     & T3                     & T4                     & T5                     & T6                     & T7                     \\ \cline{3-9} 
		& \multicolumn{1}{l|}{T1} & \multicolumn{1}{l|}{}  & \multicolumn{1}{l|}{}  & \multicolumn{1}{l|}{\cellcolor{black}}& \multicolumn{1}{l|}{}  & \multicolumn{1}{l|}{}  & \multicolumn{1}{l|}{}  & \multicolumn{1}{l|}{}  \\ \cline{3-9} 
		& \multicolumn{1}{l|}{T2} & \multicolumn{1}{l|}{\cellcolor{black}}& \multicolumn{1}{l|}{}  & \multicolumn{1}{l|}{}  & \multicolumn{1}{l|}{}  & \multicolumn{1}{l|}{}  & \multicolumn{1}{l|}{\cellcolor{black}}& \multicolumn{1}{l|}{}  \\ \cline{3-9} 
		& \multicolumn{1}{l|}{T3} & \multicolumn{1}{l|}{}  & \multicolumn{1}{l|}{}  & \multicolumn{1}{l|}{}  & \multicolumn{1}{l|}{\cellcolor{black}}& \multicolumn{1}{l|}{}  & \multicolumn{1}{l|}{}  & \multicolumn{1}{l|}{}  \\ \cline{3-9} 
		Host & \multicolumn{1}{l|}{T4} & \multicolumn{1}{l|}{}  & \multicolumn{1}{l|}{\cellcolor{black}}& \multicolumn{1}{l|}{}  & \multicolumn{1}{l|}{}  & \multicolumn{1}{l|}{}  & \multicolumn{1}{l|}{}  & \multicolumn{1}{l|}{\cellcolor{black}}\\ \cline{3-9} 
		& \multicolumn{1}{l|}{T5} & \multicolumn{1}{l|}{}  & \multicolumn{1}{l|}{}  & \multicolumn{1}{l|}{}  & \multicolumn{1}{l|}{}  & \multicolumn{1}{l|}{\cellcolor{black}}& \multicolumn{1}{l|}{}  & \multicolumn{1}{l|}{}  \\ \cline{3-9} 
		& \multicolumn{1}{l|}{T6} & \multicolumn{1}{l|}{}  & \multicolumn{1}{l|}{}  & \multicolumn{1}{l|}{\cellcolor{black}}& \multicolumn{1}{l|}{}  & \multicolumn{1}{l|}{}  & \multicolumn{1}{l|}{}  & \multicolumn{1}{l|}{}  \\ \cline{3-9} 
		& \multicolumn{1}{l|}{T7} & \multicolumn{1}{l|}{\cellcolor{black}}& \multicolumn{1}{l|}{}  & \multicolumn{1}{l|}{}  & \multicolumn{1}{l|}{}  & \multicolumn{1}{l|}{}  & \multicolumn{1}{l|}{\cellcolor{black}}& \multicolumn{1}{l|}{}  \\ \cline{3-9} 
	\end{tabular}
	\caption{Experimental designs in different cases, corresponding to two-dimensional experiment similarity chart $\mathbb{Z}^2$. Each cell is an experiment, and neighboring cells are similar experiments. Black cells are conducted experiments, and white cells are non-conducted experiments. Black cells are not neighboring. Each non-boundary white cell is neighboring to $k$ black cells, where $k=4$ (upper), $k=2$ (middle), $k=1$ (lower) \cite{wang2021inference}. This table corresponds to the contents of Subsection~\ref{s4.1}.}
	\label{co}
\end{table}

In this general setting, we need to color each node of the $n$-dimensional square lattice $\mathbb{Z}^n$ black or white, so that each white node is neighboring to at least $k$ black node, and the number of black nodes is minimized. Here each node has a coordinate $(x_1,x_2,\ldots,x_n)$, and two nodes are neighboring if one component of their coordinates differs by $1$, and other components are equal. Each white node is neighboring to at least $k$ black nodes, and each black node is neighboring to at most $2n$ white nodes. Therefore, the theoretical upper bound of white-black ratio is $2n:k$, and the minimal proportion of black nodes is $k/(2n+k)$. The following results solve the coloring problem for certain $k$ \cite{wang2022ar}:
\begin{theorem}
	If $k$ is a divisor of $n$, color a node black if and only if its coordinate satisfies $a_1x_1+a_2x_2+\cdots+a_nx_n\equiv 0 \text{ } (\text{mod }  (2n/k)+1)$, where the coefficients $a_1,a_2,\ldots,a_n$ are $k$ groups of $1,2,\ldots,n/k$: $1,2,\ldots,n/k,1,2,\ldots,n/k,\ldots,1,2,\ldots,n/k$. Then black nodes are not neighboring, each white node is neighboring to $k$ black nodes, and the proportion of black nodes is $k/(2n+k)$.
\end{theorem}
\begin{theorem}
	If $k$ is a divisor of $2n$, but not a divisor of $n$, color a node black if and only if its coordinate satisfies $a_1x_1+a_2x_2+\cdots+a_nx_n\equiv 0 \text{ } (\text{mod }  (2n/k)+1)$, where the coefficients $a_1,a_2,\ldots,a_n$ are $k/2$ groups of $1,2,\ldots,2n/k$: $1,2,\ldots,2n/k,1,2,\ldots,2n/k,\ldots,1,2,\ldots,2n/k$. Then black nodes are not neighboring, each white node is neighboring to $k$ black nodes, and the proportion of black nodes is $k/(2n+k)$.
	\label{prop2}
\end{theorem}
This is another example that cancer research can lead to mathematical progresses.

\subsection{Other topics in developmental biology}
When transplanting cancer cells into an embryo, the position of transplantation determines the fates of these cancer cells \cite{kagawa2022human}. This leads to a central problem in developmental biology: why cells at different positions of an embryo know where they are, and differentiate accordingly into normal tissues? The notion of positional information is invented to explain this problem \cite{wolpert2016positional}. Nevertheless, the usages of this term cause confusions, since it might mean different objects in different studies. In a recent work, the definition of positional information is clarified with some criteria: it is position-related; it is extracellular; it is the direct cause of cell fate \cite{wang2020biological}. This method of clarifying a notion with criteria can be applied to other fields \cite{wang2022impossibility}.

To study the development of an embryo, one convenient way is to represent the development process as a rooted tree, where each node is a cell along with its cell event (label), while an edge links the parent cell and the child cell. Besides, we cannot distinguish two children cells after division. Therefore, each development process is a rooted unordered tree with possibly repeated labels. To compare such trees, there are different metrics, where some can be calculated within polynomial time \cite{wang2022two}. One straightforward idea is to calculate the minimal number of operations to transform one tree into another. Another idea is to transform both trees into some regular forms and compare them.

\section{Discussion}
In this review, we go through three facets of mathematical cancer biology: population dynamics, gene regulation, and developmental biology. These fields have many meaningful mathematical problems, and solving them can help with the understanding of cancer. Besides, skills in mathematical modeling and data analysis are also essential. There are other fields related to cancer biology that need applied mathematicians. Although most mathematical results in this review are related to probability and discrete mathematics, other mathematical tools might be useful in such fields, such as differential equations and dynamical system, and even geometry. We hope that this review can inspire more applied mathematicians to work on mathematical problems related to cancer biology.

\bibliographystyle{acm}
\bibliography{cr}

\end{document}